# The acceleration of electrons in Radio Supernova SN1986J

Lewis Ball[1] and J.G. Kirk[2]

[1] Research Centre for Theoretical Astrophysics, University of Sydney, NSW 2006, Australia
[2] Max-Planck-Institut für Kernphysik, Postfach 10 39 80, Heidelberg D-69029, Germany

**Abstract.** We propose a model for radio supernovae (RSN) based on the synchrotron emission from relativistic electrons which are diffusively accelerated at the expanding supernova shock. This model was originally developed for application to the optically thin emission observed from SN1987A. Here we generalise it by including the effects of free-free absorption from both an external screen and from material internal to the source, and by relaxing the restriction to an azimuthal $B$-field. We find a good fit to the entire set of radio data for the best observed highly-luminous RSN – SN1986J – with a reduced $\chi^2$ of 3.85. Applying the new model to SN1988Z, another intrinsically bright RSN, also yields a good fit ($\chi^2_{\rm red} \approx 2$) but this is less significant, because of the limited data on this distant ($z = 0.02$) source. These fits suggest that the shock expands at constant speed, that the magnetic field within the source decreases with time according to $t^{-2}$, and that the compression ratio of the shock front is close to the value expected of a strong shock in an ideal gas of adiabatic index 5/3 – indicating a relatively low value of the cosmic ray pressure compared with SN1987A. In the case of SN1986J we derive an explosion date in August/September 1982, a magnetic field at the position of the shock 1000 days after explosion of $B \approx 4\,{\rm nT}$ and a spatial diffusion coefficient of the electrons of $\kappa \approx 4 \times 10^{19}\,{\rm m}^2\,{\rm s}^{-1}$, four orders of magnitude greater than the Bohm value. In addition, we obtain the optical depths to external and internal absorption, and derive an estimate of the mass-loss rate.

## 1. Introduction

All radio supernovae (RSN) detected so far show a pattern typical for a source which is initially optically thick and later becomes optically thin: the emission appears first at high frequencies and extends later to lower ones. At fixed frequency the flux rises, reaches a peak and then fades somewhat more slowly than it rose. Once the radio emission is well past its peak, the flux density decreases with increasing frequency and is well-fitted by a power law of spectral index between 0.5 and 1.2 – a strong indication that it is synchrotron radiation from

*Send offprint requests to:* J.G. Kirk

relativistic electrons. This behaviour indicates that the radio emission originates behind the outwardly propagating shock front. At early epochs the ionised component of the stellar wind of the progenitor upstream of the shock front is sufficiently dense for free-free absorption by thermal electrons to prevent the escape of the synchrotron radiation. As the shock propagates out through the overlying wind material the effective optical depth decreases allowing the radiation to escape (Chevalier 1982; Weiler et al. 1986).

However, the observed flux results from a convolution of the absorption process with the process responsible for the intrinsic emission, which itself depends on the way in which electrons are accelerated and on the evolution of the magnetic field within the source. Chevalier (1982) suggested that the intrinsic radio emission from RSN could be understood by assuming that a fixed fraction (about 2%) of the flux of energy through the shock front is divided equally between relativistic electrons and magnetic field. Approximating the emission region as a thin shell behind the shock, and including the effect of free-free absorption in a stellar wind with density falling off with the square of the distance from the progenitor star, Chevalier predicted that the radio flux density $S(\nu, t)$ as a function of frequency $\nu$ and time $t$ should be described by a five parameter model. Weiler et al. (1986) proposed a more general empirical model with six free parameters, of which the Chevalier (1982) model is a special case. The data from several Type Ib and Type II RSN (most notably SN1979C and SN1980K) were shown by Weiler et al. (1986) to be well-fitted by both the general model and Chevalier's model. However, these models failed to fit one of the brightest and best-observed RSN: SN1986J.

First discovered as a radio source, SN1986J was subsequently found to have prediscovery radio and optical detections. The explosion date remains uncertain; optical spectra indicate that it was a "peculiar" Type II supernova, and it is thought to have involved the explosion in 1982–83 of an extremely massive progenitor with perhaps $M \sim 20 - 60\,{\rm M}_\odot$ (Rupen et al. 1987; Weiler & Sramek 1988). The radio emission from SN1986J rose much more slowly than is compatible with the rather rapid drop in absorption implied in the standard model. This led to the suggestion that there might be an additional contribution to the free-free absorption due to thermal material which is mixed with the nonthermal synchrotron-emitting material (Weiler et al. 1990 – henceforth "WPS"), an effect which is easily incorporated for a slab-like geometry giv-ing a significantly better fit to the data from early epochs.

In previous work (Ball & Kirk 1992 – henceforth "BK") we developed a model for the unique second phase of radio emission from SN1987A (Staveley-Smith et al. 1992, 1993; Ball et al. 1995) based on the diffusive acceleration of electrons at the expanding supernova shock. It has since emerged that the emission from SN1987A requires for its understanding a detailed treatment of the environment of the progenitor and the reaction of accelerated particles on the hydrodynamics of the explosion (Duffy et al. 1995). In this *Letter* we suggest that in the case of SN1986J such complications can be neglected, and show that it is possible to fit a simple form of the model (including a combination of internal and external free-free absorption as proposed by WPS) to the radio data. This has the advantage that the free parameters are simply related to physical properties of the explosion, enabling estimation of quantities such as the compression ratio of the shock front, the magnetic field and the effective diffusion coefficient of the electrons.

In Sect. 2, we describe the basic properties of the diffusive shock acceleration model. We identify the free parameters of the model and discuss their relation to the physical parameters describing the shock and the environment into which it expands. In Sect. 3 we present the results of using a $\chi^2$-minimisation procedure to fit the model to the long term evolution of the radio emission from SN1986J. In Sect. 4 we discuss this result and its implications, and note that fits to the observations of SN1988Z, another peculiar Type II supernova, lend further support to the diffusive acceleration model.

## 2. The model

The model presented by BK assumes diffusive particle acceleration at a shock expanding at constant speed so that the shock radius $r_s$ is proportional to the time after explosion, $r_s \propto (\hat{t} - \hat{t}_e)$ where $\hat{t} = t/(1\,\mathrm{day})$ is the dimensionless time variable and $\hat{t}_e$ is the time of the explosion. This is the behaviour expected in the similarity solutions found by Chevalier (1982) if the density profile of the ejecta is very steep. It is also consistent with almost all of the published fits to the radio emission of other RSN (Weiler et al. 1986). In the case of SN1987A, the magnetic field in the wind of the progenitor was assumed to be predominantly azimuthal. This leads to a time dependence of the magnetic field at the shock front $B(r_s)$ of the form $B(r_s) \propto (\hat{t} - \hat{t}_e)^{-1}$. If, on the other hand, the magnetic field encountered by the shock is predominantly radial, then $B(r_s) \propto (\hat{t} - \hat{t}_e)^{-2}$. Here we allow for each of these situations by setting $B(r_s) \propto (\hat{t} - \hat{t}_e)^{-n}$, where $n$ is a free parameter. The model also assumes that the rate of injection of electrons into the acceleration mechanism is a constant: given that the shock speed does not vary, this corresponds to the injection of a fixed fraction of the electrons from a wind of density varying as the inverse square of the distance $r$ from the progenitor. The resulting intrinsic (unabsorbed) flux density can be written as a function of the dimensionless frequency $\hat{\nu} = \nu/(5\,\mathrm{GHz})$ and time in the form

$$F(\hat{\nu}, \hat{t}) = K_1\, \hat{\nu}^{-\alpha}\, (\hat{t} - \hat{t}_e)^{1-n(1+\alpha)}\, \psi(\hat{\nu}, \hat{t}, \alpha, \hat{t}_e, n, \hat{t}_{\mathrm{acc}}, \nu_0) \quad (1)$$

where $\psi(\hat{\nu}, \hat{t}, \alpha, \hat{t}_e, n, \hat{t}_{\mathrm{acc}}, \nu_0)$ is a well-behaved function, which is constant at large $\hat{t}$, and which can be found from a straightforward generalization of Eq. (8) of BK:

$$\psi(\hat{\nu}, \hat{t}, \alpha, \hat{t}_e, n, \hat{t}_{\mathrm{acc}}, \nu_0) \equiv$$

$$\tau_0^{1-n(1+\alpha)} \int_{\tau_0}^{\tau_1} d\tau\, \tau^{-2+n(1+\alpha)} (1-\tau)^{4\alpha/3} \,+$$

$$\frac{\hat{t}_{\mathrm{acc}}}{2\alpha \hat{t}} \left( \frac{3}{2\alpha \tau_0} \right)^{-1-4\alpha/3} \quad (2)$$

where $\tau_0 = 3/(3 + 2\alpha)$ and $\tau_1 = \{1 + R_1/[\tau_0(\hat{t} - \hat{t}_e)]\}^{-1}$, with $R_1$ given by the solution of the transcendental equation

$$\nu \left[ R_1 + \tau_0(\hat{t} - \hat{t}_e) \right]^{(4+3n)/3} [3R_1/(2\tau_0\alpha)]^{-4/3} =$$

$$\nu_0 \exp[3R_1/(\tau_0 \alpha \hat{t}_{\mathrm{acc}})]\, . \quad (3)$$

[Note that the integral in Eq. (2) can be expressed in terms of incomplete Beta functions (BK).] This model for the intrinsic flux density has six free parameters all of which can be directly identified with physical properties of the source and/or acceleration mechanism. They are: (i) the overall flux scaling parameter $K_1$; (ii) the spectral index $\alpha$ which gives the shock compression ratio $\rho_c = (3 + 2\alpha)/(2\alpha)$; (iii) the epoch of explosion $\hat{t}_e$; (iv) the index $n$ giving the time dependence of the magnetic field strength $B \propto (\hat{t} - \hat{t}_e)^{-n}$; (v) the acceleration time $\hat{t}_{\mathrm{acc}}$, which is related directly to the electron diffusion coefficient $\kappa$ and the compression ratio, provided the shock speed $v_s$ is known: $\kappa = t_{\mathrm{acc}} v_s^2 (\rho_c - 1)/[3\rho_c(\rho_c + 1)]$; (vi) the frequency scaling parameter $\nu_0$, which is related to the magnetic field strength at the shock front and the momentum $p_0$ with which electrons are injected: $\nu_0 = 10^{3n} a_1 (p_0/m_e c)^2 B_{1000}$, where $B_{1000}$ is the magnetic field at the shock front at age 1000 days, and $a_1 = 1.3 \times 10^{10}\,\mathrm{Hz\,Tesla}^{-1}$.

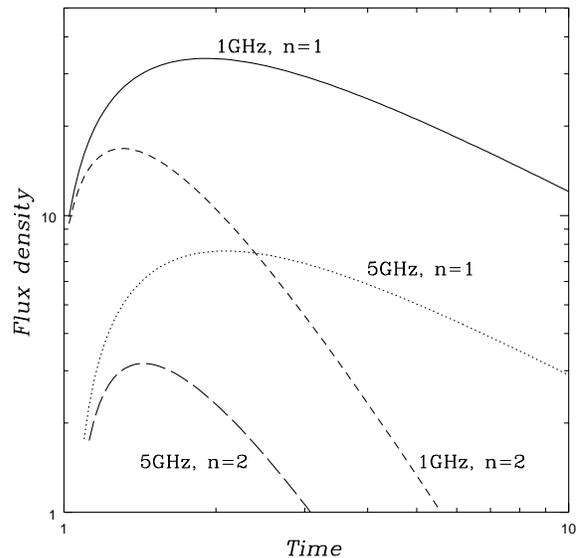

**Fig. 1.** The intrinsic emission from the diffusive acceleration model at two frequencies differing by a factor of five – 1 GHz and 5 GHz say – for $n = 1$ and $n = 2$.

Fig. 1 shows the intrinsic flux density as given by the diffusive acceleration model at two frequencies differing by a factor of five, for the cases $n = 1$ and $n = 2$. The plots clearly show the initial increase in the intrinsic flux density at a given frequency, the fact that the intrinsic emission looks "optically thin" at all times (i.e. the intrinsic flux density always decreases with increasing frequency), and that the eventual decay is a power-law in time which is steeper when $n$ is larger.

The BK model for the second phase of radio emission from SN1987A does not include absorption effects because the onset of the emission was optically thin, indicating that by then the shock had expanded sufficiently for the absorption to be negligible. This is not the case for the emission observed from other RSN so that the effects of absorption must be included. We treat free-free absorption in the same way as WPS, so that the radio flux density is given by

$$S(\hat{\nu}, \hat{t}) = A_e(\tau) A_i(\tau') F(\hat{\nu}, \hat{t}) \qquad (4)$$

where $F(\hat{\nu}, \hat{t})$ is the intrinsic (unabsorbed) emission,

$$A_e(\tau) = e^{-\tau} \quad \text{with} \quad \tau = K_2 \hat{\nu}^{-2.1} (\hat{t} - \hat{t}_e)^{-3} \qquad (5)$$

describes the external absorption by the wind, and

$$A_i(\tau') = (1 - e^{-\tau'})/\tau' \quad \text{with} \quad \tau' = K_3 \hat{\nu}^{-2.1} (\hat{t} - \hat{t}_e)^{-5} \qquad (6)$$

describes the 'internal' absorption by material which is mixed with (at least in a statistical sense) the nonthermal radiating electrons. These formulae apply for a shock moving at constant speed. (Note that Eqs. (5) and (6) imply in the notation of WPS $\delta = -3$ and $\delta' = -5$.)

External absorption is related to the density of the environment and, hence, to the mass-loss rate of the progenitor. The parameter $K_2$ is determined by the density of the progenitor's wind, which is assumed proportional to $r^{-2}$, and from Weiler et al. (1986), we find

$$\dot{M} = 10^{-8} K_2^{1/2} w_{10} v_{s,4}^{3/2} T_4^{0.68} \, M_\odot \, \text{yr}^{-1} , \qquad (7)$$

where $w_{10} \times 10 \, \text{km s}^{-1}$ is the speed of the progenitor's wind, $T_4 \times 10^4$ K its temperature (after explosion) and $v_s = v_{s,4} \times 10^4 \, \text{km s}^{-1}$.

Internal absorption can be interpreted as due to either a homogeneous mixture of emitting and absorbing plasma, or to a statistical superposition of emitting and absorbing clumps along the line of sight. The parameter $K_3$ is determined by the total number of thermal electrons mixed with the radiating material, which is assumed constant. The relevant physical parameter is the effective optical depth $\tau'_{1000}(5\,\text{GHz})$ at 1000 days after explosion, measured at a frequency of 5 GHz: $\tau'_{1000}(5\,\text{GHz}) = 10^{-15} K_3$.

In total, the model we propose has eight free parameters: six for the intrinsic flux density and two more, $K_2$ and $K_3$, for external and internal absorption. At early times the evolution of the flux density and the frequency spectrum are determined by a combination of the increasing intrinsic emission and the decreasing absorption. At large times, when absorption is negligible, the diffusive model gives a flux density $S(\hat{\nu}, \hat{t}) \propto \hat{\nu}^{-\alpha} (\hat{t} - \hat{t}_e)^{1-n(1+\alpha)}$ so that for a given value of the spectral index $\alpha$ the flux decline at late times is faster when $n$ is larger. (Note also that the asymptotic behaviour of the diffusive model is equivalent to that of the WPS model if in the notation of WPS one identifies $\beta = 1 - n(1+\alpha)$.)

## 3. Fits to the observations

We have used the $\chi^2$-minimization technique to find the best fit of the diffusive acceleration model with both external and internal absorption, as given by Eqs. (1–6), to the observed emission from RSN. Using four-frequency observations of the radio emission from SN1986J extending to 1991 June 11 (WPS;

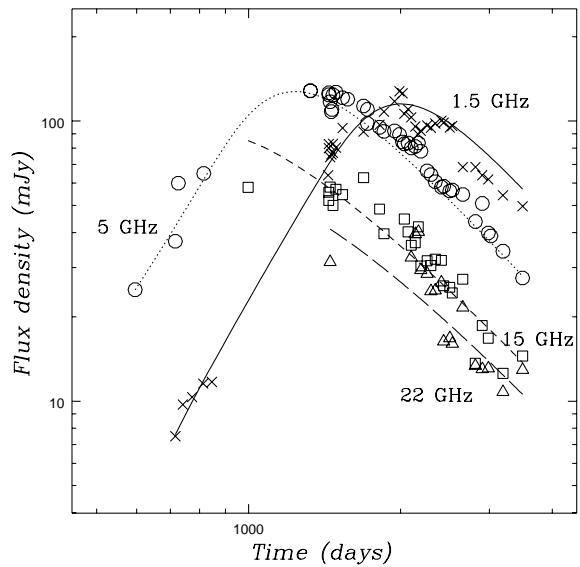

**Fig. 2.** Best fit of the diffusive acceleration model (curves) to the observed radio emission from SN1986J up to 1991 June 11. Crosses indicate measurements at 1.5 GHz; circles 5 GHz; squares 15 GHz and triangles 22 GHz.

Weiler, van Dyk & Sramek, private communication) we obtain an acceptable fit with $\chi^2$ per degree of freedom $\chi^2_{\text{red}} = 3.85$, where $\chi^2_{\text{red}} = \chi^2/(n_d - n_p)$ with $n_d$ the number of data points and $n_p$ the number of free model parameters. The parameters of this best fit model are given in Table 1, and Fig. 2 shows the resulting flux densities together with the data.

The value of $\chi^2_{\text{red}}$ for the best fit diffusive model to the SN1986J data is comparable to that obtained for model fits of the Chevalier mini-shell model by WPS (to observations extending to 1988 December 28). The model parameters that are a direct analogue of those of the model fits by WPS are $\alpha$, $\hat{t}_e$, $K_2$ and $K_3$. The value of $\alpha$ that we obtain ($\alpha = 0.57$) is somewhat smaller than that found by WPS ($\alpha = 0.67$). This difference arises at least partly because the extended data set now available contains more information on the later optically thin stages of the emission. The value of $\hat{t}_e$ quoted in Table 1 is expressed relative to the best fit explosion time found by WPS, and is consistent with their fit. The values of $K_2$ and $K_3$ that we obtain are both considerably larger than the values given by the best fit model quoted by WPS, but differ somewhat in their definition, since WPS find a decelerating shock front.

A fit of the diffusive acceleration model to the luminous radio supernova SN1988Z (see van Dyk et al. 1993 for a detailed discussion of the properties of this source), also gives an acceptable $\chi^2_{\text{red}} \approx 2$. The best-fit model implies that in SN1988Z, as in SN1986J, the magnetic field at the shock falls off with the inverse square of the time since explosion ($B \propto (\hat{t} - \hat{t}_e)^{-2}$), and that the shock front is almost unmodified ($\rho_c \approx 4$). However, the relatively small number of observations of this distant source, and their necessarily large associated uncertainties, mean that the results of model fits to SN1988Z are less significant than those to SN1986J. Nevertheless, these sources are both peculiar Type II supernovae, which are qualitatively similar, and it is encouraging that the fits of the diffusive ac-celeration model imply that the physical source parameters are also similar.

**Table 1.** Parameters for the best fit of the diffusive acceleration model to the observed radio emission from SN 1986 J. No uncertainties are quoted for $K_1$, $K_2$ and $K_3$ because they are composites of the free parameters which were minimised.

| Parameter | | Physical interpretation |
|---|---|---|
| $K_1$ | $1.8 \times 10^{10}$ | |
| $K_2$ | $1.1 \times 10^{7}$ | $\dot{M} = 3.4 \times 10^{-5} w_{10} v_{s,4}^{3/2} T_4^{0.68} \, M_\odot \, \mathrm{yr}^{-1}$ |
| $K_3$ | $2.5 \times 10^{15}$ | $\tau'_{1000}(5\,\mathrm{GHz}) = 2.5$ (internal absorption) |
| $\alpha$ | $0.58 \pm 0.01$ | $\rho_c = 3.61 \pm 0.02$ |
| $\hat{t}_e$ | $-21 \pm 26$ | $t_e = 1982$ Aug. 24 |
| $n$ | $2.01 \pm 0.06$ | $B \propto t^{-2}$ |
| $\hat{t}_{\mathrm{acc}}$ | $332 \pm 25$ | $\kappa = 4.2 \times 10^{19} v_{s,4}^2 \, \mathrm{m}^2 \, \mathrm{s}^{-1}$ |
| $\nu_0/10^{13}$ | $4.8 \pm 2.2$ | $B_{1000}\,(p_0/m_e c)^2 = 3.5 \pm 1.7 \, \mathrm{nT}$ |
| $\chi^2_{\mathrm{red}}$ | $3.85$ | |

## 4. Discussion and Conclusions

The salient points of the best-fit diffusive model to the radio-luminous SN 1986J and SN 1988Z are (i) a magnetic field dependence $B \propto t^{-2}$ and (ii) a compression ratio not too different from the value 4 expected at a shock front which is not modified by accelerated particle pressure. A magnetic field dependence of this type could arise if the field in the progenitor's wind were close to radial – perhaps as a result of a very small rotation rate of the star, or of turbulent processes within the wind itself. It is interesting to note that such a rapid decrease in the field at the position of the shock front is more difficult to achieve if the shock suffers appreciable deceleration, since in this case even a radial field would weaken more slowly than $t^{-2}$. The value of the compression ratio indicates that the shock front has been slightly modified by the pressure of accelerated particles, but that this has not been as effective as in the case of SN 1987A (Duffy et al. 1995). Such an effect may be connected with the sharp fall-off of the magnetic field. The pressure in relativistic particles is dominated by particles of TeV energy or higher, and a weak magnetic field will dramatically increase the time needed to accelerate particles to such energies. The magnetic field at the shock position 1000 days after explosion is found to be about 3 nT, significantly stronger than the ambient interstellar field. This is consistent with our picture, in which the field is related to the progenitor. Extrapolating back to the stellar surface implies a prohibitively high surface field, but the model does not require the $t^{-2}$ behaviour to apply earlier than about 500 days after explosion. The effective diffusion coefficient we derive is equivalent to a mean free path of an electron of $10^{12}$ m, some four orders of magnitude greater than the gyro-radius of a radio emitting electron. This finding is similar to that for SN 1987A (BK) although, in the present case, the magnetic field geometry does not require particles to propagate normal to the direction of the field. Our value of the mass-loss rate of the progenitor is somewhat larger than found by WPS, and the optical depth we find for free-free absorption by material mixed with the emitting source is slightly higher at 1000 days. However, as noted by WPS, the errors, especially in the estimate of the mass-loss rate, are large.

The diffusive acceleration model also provides adequate fits to the radio emission from other less luminous but well-observed RSNe, namely the Type Ib RSN SN1983N, and the "normal" Type II RSNe SN1979C and SN1980K. In contrast to SN1986J and SN1988Z, the fits to these other RSN imply a magnetic field dependence $B \propto t^{-1}$ which is consistent with the modelling of the second phase of emission from SN1987A (BK). This points to a significant difference in the progenitors of different supernova types. There are also indications that the emission from the recent RSN SN1993J requires a more detailed approach to the modelling, and in particular that the density of the circumstellar wind into which the shock is expanding decreases in this object more slowly than $r^{-2}$ (van Dyk et al. 1994; Suzuki & Nomoto 1995). Thus, differences between the models for the increasing number of well-observed RSN may well provide useful insights into the properties of these sources, and help provide a physical basis for the various supernova classifications.

The diffusive model has more free parameters (a total of eight) than the Chevalier mini-shell model and the more general empirical models previously used to interpret the observations of RSN. However, the value of $\chi^2$ per degree of freedom provides an objective measure of the quality of the fit, independent of the number of free parameters, and the best fits of the diffusive acceleration model are comparable in quality to the fits of the other models to SN1986J (WPS), SN1988Z (van Dyk et al. 1993) and to less luminous RSN such as SN1979C (Weiler et al. 1986). Furthermore, the proposed model has the advantage that it involves a specific physical mechanism for accelerating electrons to the relativistic energies required to produce the observed radio emission.

*Acknowledgements.* We thank K. Weiler, S. D. van Dyk and R. A. Sramek for providing us with unpublished data, P. Duffy and K. Weiler for useful discussions and the Max-Planck-Gesellschaft for the grant of a visiting fellowship to one of us (LB).